\documentclass[letterpaper,11pt]{article}
\pdfoutput=1 % if your are submitting a pdflatex (i.e. if you have
\usepackage{jheppub} % for details on the use of the package, please
\usepackage{dsfont}
\input epsf
\newcommand{\vev}[1]{\langle #1 \rangle}

\usepackage{caption}
\usepackage{subcaption}	

\def\Tr{{{\rm Tr~ }}}

\def\vev#1{\langle{#1}\rangle}

\definecolor{dgreen}{rgb}{0.0, 0.42, 0.24}

\def\T{{\cal T}}

\title{Path Integral Complexity and Kasner singularities }

\author{Pawel Caputa$^1$,}
\author{Diptarka Das$^2$,}
\author{Sumit R. Das$^3$,}

\affiliation{$^1$ Faculty of Physics, University of Warsaw, 02-093 Warsaw, POLAND.}
\affiliation{$^2$ Department of Physics, Indian Institute of Technology, Kanpur, UP 208016, INDIA.}
\affiliation{$^3$Department of Physics and Astronomy, University of Kentucky, Lexington, KY 40506, U.S.A.}

\emailAdd{Pawel.Caputa@fuw.edu.pl}\emailAdd{didas@iitk.ac.in}\emailAdd{das@pa.uky.edu}

\abstract{We explore properties of path integral complexity in field theories on time dependent backgrounds using its dual description in terms of Hartle-Hawking wavefunctions. In particular, we consider boundary theories with time dependent couplings which are dual to Kasner-AdS metrics in the bulk with a time dependent dilaton. We show that holographic path integral complexity decreases as we approach the singularity, consistent with earlier results from holographic complexity conjectures. Furthermore, we find examples where the complexity becomes universal i.e., independent of the Kasner exponents, but the properties of the path integral tensor networks depend sensitively on this data.}

\begin{document}
\maketitle
\flushbottom

\def\ben{\begin{equation}}
\def\een{\end{equation}}
\def\bea{\begin{eqnarray}}
\def\eea{\end{eqnarray}}
\def\vx{\vec{x}}
\def\vn{\vec{n}}
\def\teta{\tilde{\eta}}
\def\tg{\tilde{g}}
\def\tb{\tilde{b}}
\def\tbeta{\tilde{\beta}}
\def\cC{{\cal{C}}}
\def\cb{\color{blue}}
\newcommand{\cg}{\color{dgreen}}

%%%%%%%%%%%%%%%%%%%%%%%%%%%%%%%%%%%%%%%%%%%%%%%%%%%%%%%%
%%%%%%%%%%%%%%%%%%%%%%%%%%%%%%%%%%%%%%%%%%%%%%%%%%%%%%%%
\section{Introduction and Summary}
\label{sec:intro}
%%%%%%%%%%%%%%%%%%%%%%%%%%%%%%%%%%%%%%%%%%%%%%%%%%%%%%%%
%%%%%%%%%%%%%%%%%%%%%%%%%%%%%%%%%%%%%%%%%%%%%%%%%%%%%%%%
In the past several years there has been a lot of progress in understanding the role of complexity in the AdS/CFT correspondence. In \cite{susskind1}-\cite{susskind3} it was noted that the complexity of a state needs to be an essential ingredient in the emergence of a smooth gravitational dual, and holographic complexity formulae in AdS were proposed \cite{holocomplexreview}. These conjectures have by now been studied intensely in various static and dynamic geometries (see e.g. review \cite{Chapman:2021jbh}). On the field theory side, various notions of complexity have been explored and their relations to each other and to holographic notions have been discussed (see e.g. \cite{caputa-taka,Czech:2017ryf,ComplexityQFT,Caputa:2018kdj,Magan:2018nmu,Chagnet:2021uvi} and references therein).

A particularly useful notion of state complexity was proposed in \cite{caputa-taka}. The idea is that the complexity of a state in a field theory is related to the optimal way of preparing the wavefunction with a path integral. Consider for example the ground state of a field theory on flat Euclidean space. The vacuum wavefunctional on some time slice can be computed by calculating the Feynman path integral on an infinite strip which starts from infinitely large negative Euclidean time and ends on the time slice of interest. As it is often the case, such a calculation can be performed by replacing the base space by a lattice and taking the continuum limit appropriately. However, at large Euclidean times, long wavelength modes dominate this path integral - the dominant range of wavelengths decrease as one approaches the time slice in question. It is therefore much more efficient to calculate this path integral on a lattice with a variable lattice spacing, with the lattice spacing decreasing as one approaches the time slice where the boundary conditions are specified - provided the dependence on these boundary conditions (which are the arguments of the wavefunctional) is not changed in this  modified calculation. This means that the modified path integral should be proportional to the original one and the proportionality factor should not depend on the boundary values of the fields. Based on this intuition, the Path Integral Optimization \cite{caputa-taka} then amounts to minimizing this proportionality factor that is defined as the exponent of the Path Integral Complexity action.

In the continuum, this procedure is equivalent to replacing the original flat base space by a curved geometry, e.g. by a metric which is conformal to flat space. For two dimensional conformal field theories (CFT) the proportionality factor becomes the exponential of the Liouville action. Thus, in the optimization, we need to find solutions of the Liouville equations with appropriate boundary conditions. The on-shell Liouville action evaluated on these solutions then defines a notion of complexity, which is called Path Integral Complexity.\footnote{See e.g. \cite{Czech:2017ryf,Caputa:2018kdj,complex-circuit-pio,Erdmenger:2020sup}  for discussions relating the path integral complexity to other, more standard notions of complexity.}. Moreover, it was found in \cite{caputa-taka} that the optimal metrics for various states in 2d CFTs are in fact hyperbolic - they were then interpreted as the induced metrics on a constant time slices in the dual $AdS_3$ geometry. This can be seen as an explicit realization of tensor network (TN) constructions of holography \cite{AdSTN,Caputa:2020fbc}. In \cite{bcdkmt} this construction was extended to two dimensional CFT's deformed by relevant operators. The (perturbative) optimization procedure then requires not only a nontrivial base space metric, but also a Euclidean time dependent operator coupling in a way which is consistent with the renormalization group. The optimal metric is again consistent with a time slice of the  corresponding bulk dual geometry. When the original base space of the CFT is curved \cite{Caputa:2020mgb} or contains boundaries \cite{Sato:2019kik}, this procedure still works and leads to a consistent picture.

More recently, \cite{HHcomplex} proposed a holographic dual to the path integral optimization procedure. The idea is to evaluate the Hartle-Hawking wavefunctional in $AdS$ spacetime using gravity path integral in a region $M$ from the asymptotic cutoff boundary $\Sigma$ up to a co-dimension one surface $Q$ that intersects $\Sigma$ at some time and extends into the bulk. The AdS/CFT correspondence can then be used to identify this quantity with a wavefunctional of the field theory on the boundary. In addition, the surface $Q$ is endowed with a tension $T$ introduced by a corresponding boundary term added to the Einstein-Hilbert action. In the semiclassical limit, the wavefunction is approximately evaluated by saddle point, and the on-shell action is then maximized with respect to variations of the surface $Q$. It has been argued in \cite{HHcomplex} that for a non-zero $T$, this procedure corresponds to a  partial path integral optimization of the boundary theory, whereas for $T=0$ one gets a full optimization. The on-shell gravity action then provides a notion of {\em Holographic Path Integral Complexity} (HPIC). It was shown that this procedure agrees with path integral optimizations studied in \cite{caputa-taka} for various states ranging from the vacuum to interesting excited states including thermal states. This scheme naturally extends to arbitrary number of dimensions and provides a way to extend the discussion to the (bulk) quantum regime.
Significantly, this Hartle-Hawking procedure continues to work in Lorentzian signature and yields reasonable results for de Sitter and Anti-de-Sitter bulks.

In this paper we take the first step in extending the discussion to non-trivial dynamical spacetimes and conformal field theories on time dependent backgrounds. In particular, we will concentrate on CFT on cosmological backgrounds which generically behave as Kasner-type metrics near singularities as well as time dependent couplings, and consider their bulk duals which include a time dependent dilaton.
The study of such backgrounds was initiated in \cite{dmnt} for a five dimensional bulk and continued in \cite{adnt}-\cite{brand}. It was shown that  any solution of four dimensional dilaton gravity with vanishing cosmological constant can be lifted to a solution of five dimensional dilaton gravity with a negative cosmological constant. The five dimensional bulk solution can be then regarded as the gravity dual of a four dimensional CFT living on a nontrivial metric and with a space-time dependent coupling constant. This led to a large class of solutions which depend only on either time or a null coordinate. This class includes four dimensional Kasner metrics with nontrivial dilatons \footnote{Such solutions have been studied in the purely four dimensional context many years ago \cite{misner}.}. The solutions with null singularities were also discussed in \cite{chu}. The aim of these works was to explore if the boundary dual can ``resolve'' the bulk singularities. Somewhat different approaches to this question have been explored in \cite{hertog}, \cite{craps}. Kasner solutions have also been revisited more recently in \cite{engelhardt}. Holographic complexity in such bulk metric have been also studied in \cite{rabinovici}. These backgrounds also appear in a rather different recent context in \cite{hartnoll}.

It turns out that except for null solutions \cite{null} and a class of slowly varying solutions \cite{slow}, there is no unambigious answer to the question whether the dual field theory provides a smooth time evolution across the bulk singularity, and even in these cases the nature of the bulk space-time is not clear - in fact there is evidence that  for regulated solutions, the final state can be possibly a black hole \cite{slow,engelhardt}.\footnote{Nevertheless an interesting aspect of these regulated solutions is that the spectrum of cosmological fluctuations as predicted by the boundary theory do not change as one goes across the region of high curvature \cite{brand}.} We will have nothing new to say about these questions. Rather we will use Kasner backgrounds to explore boundary field theory signatures of cosmological singularities, pretty much in the spirit of \cite{engelhardt} and \cite{rabinovici}.

In the following, we will generalize these Kasner solutions with non-trivial dilaton to arbitrary number of dimensions.
These solutions are characterized by a set of Kasner exponents $p_a$ and a parameter $\alpha$ which characterizes the dilaton profile. The bulk equations of motion then lead to two constraints on the exponents, relating them to $\alpha$. Thus in $d \geq 3$ infnite classes of these exponents are possible. We then set up the calculation of Hartle-Hawking wavefunctions in these backgrounds and derive the equation for the co-dimension one surface $Q$ that extremizes the wavefunction while respecting the invariance under spatial translations on the boundary. The surface $Q$ intersects the cutoff boundary $\Sigma$ at boundary time $t_0$. For time independent backgrounds, nothing depends on $t_0$, as guaranteed by time translation invariance. However, in time dependent scenarios the Hartle-Hawking wavefunction explicitly depends on $t_0$ and we will study how HPIC depends on this parameter.

One interesting aspect of our results is that, for homogeneous $Q$, the on-shell dilaton gravity action which provides a semiclassical evaluation of the Hartle-Hawking wave function is independent of the Kasner exponents $p_a$ and independent of the dilaton profile, so long as the bulk equations of motion are satisfied. Our final results for the complexity are therefore universal within this class of backgrounds. In general, we are unable to solve the optimization equations analytically except for $d=2$ and $d=3$. For $d=2$ the boundary metric is a Milne universe, which is locally diffeomorphic to Minkowski space. In addition there is no non-trivial bulk dilaton. However, if one of the directions are made compact, these coordinate transformations are invalid and the boundary space-times are physically different, containing topological big-crunch, big-bang singularities.
The wavefunctional obtained by performing the bulk path integral as described above is not the wavefunctional in the Minkowski vacuum, but rather in a general time-dependent state whose precise nature is not known. For $d=3$ the backgrounds in the presence of a non-trivial bulk dilaton are locally different from Minkowski space. These solutions are characterized by a parameter contained in the dilaton background, and one can have an infinite set of Kasner exponents. When these exponents are not $(0,1)$ there is a genuine curvature singularity.

As in \cite{HHcomplex} we evaluate the path integral using a naive saddle point approximation and interpret the on-shell action as the HPIC of this state, which is now time dependent because of the dependence on $t_0$. We find that for both of these cases, this (relative) complexity becomes large and negative at the (cutoff) singularity, and increases as one goes further from the singularity. As we will see below the behaviour is not monotonic. Both near the singularity and far from it the complexity increases. However, when the tension $T$ exceeds a critical value there is a period of time during which the complexity decreases. This is one of our main physical results. A similar non-monotonic behaviour of complexity at early times, is known to occur due to the \emph{switchback effect} \cite{stankind} and we hope  our results will will serve as a starting point towards better understanding of this physics using path integral approach.

Finally, we interpret holographic results from the perspective of Lorentzian path integral optimization \cite{caputa-taka} as well as geometry of path integrals and Tensor Networks \cite{Milsted:2018san}. In $d=2$, we find that our optimal slices solve Lorentzian Liouville equations i.e., have constant Ricci scalar curvature and can be interpreted as Tensor Network type circuits of a time-dependent Hamiltonian. In order to gain more understanding regarding the field theory state we also compute the entanglement entropy for the 2D CFT vacuum in the Milne background. We find that the von Neumann entropy monotonically goes to zero at the cut-off singularity. The same answer is also borne out of  holography via the relevant HRT surfaces \cite{Hubeny:2007xt}. 

%%%%%%%%%%%%%%%%%%%%%%%%%%%%%%%%%%%
%%%%%%%%%%%%%%%%%%%%%%%%%%%%%%%%%%%
\section{Review of Holographic Path Integral Optimization}
\label{sec:review}
%%%%%%%%%%%%%%%%%%%%%%%%%%%%%%%%%%%
%%%%%%%%%%%%%%%%%%%%%%%%%%%%%%%%%%%
In this section we will briefly review some relevant aspects of the holographic path integral optimization in Lorentzian bulk following \cite{HHcomplex}. We will be mainly interested in computing semiclassical Hartle-Hawking wavefunctions in AdS spacetimes. For that let us introduce the basic objects and conventions. The Einstein-Hilbert action with the Gibbons-Hawking term is
\ben
I_G = \frac{1}{2\kappa^2}\int_M d^{d+1}x~\sqrt{-g} \left( R^{d+1} - 2\Lambda \right) + \frac{1}{\kappa^2} \int_{\partial M}d^dy~\sqrt{|h|} K, 
\label{1-1}
\een
where $R^{(d+1)}$ is the Ricci scalar and we have chosen units such that the cosmological constant $\Lambda$ is given by
\ben
\Lambda = -\frac{d(d-1)}{2}.
\label{1-2}
\een
$M$ denotes the $(d+1)$ dimensional space-time with a boundary $\partial M$. We will use a coordinate system $(t,\vx,z)$ with $t$ being a time coordinate and $z, \vx$ space coordinates. The region $M$ of our interest contains two pieces of the boundary (see Fig. 1 in \cite{HHcomplex}): (i) a cutoff time-like surface  $\Sigma$ at $z=\epsilon$ and (ii) a co-dimension-1 surface $Q$ which may be specified by an equation
\ben
z = f(t,\vx).
\label{1-4}
\een
This surface can be space-like or time-like.
The Gibbons-Hawking term involves an integration over $\Sigma$ and $Q$.
The induced metric on $\partial M$ is $h_{ij}(t,\vx)$, whose components can be obtained from the defining equation (\ref{1-4}). $K$ stands for the trace on the extrinsic curvature for the relevant surface. 

In the following we will consider surfaces $Q$ which are translation invariant in the $\vx$ direction so that $z=f(t)$. The surfaces $Q$ and $\Sigma$ meet at a time slice $t=t_0$ along the co-dimension-2 surface $\gamma$ with a sharp corner. We will denote the induced metric on $\gamma$ by $\gamma_{ab}$. The normal vectors $\vn_\Sigma$ and $\vn_Q$ at this intersection obey
\bea
\vn_\Sigma \cdot \vn_Q, & = & - \cosh \eta_0,~~~~~~~~~~{\rm timelike}~Q,\\
\vn_\Sigma \cdot \vn_Q, & = & - \sinh \eta_0,~~~~~~~~~~{\rm spacelike}~Q. 
\label{1-5}
\eea
To evaluate Hartle-Hawking wavefunctions we will also need a modified Hawyard term\footnote{In the notation used in \cite{HHcomplex} for spacelike $Q$ the boost parameter was written as $\tilde{ \eta}_0$. See Fig. 5 in  \cite{HHcomplex}.} \cite{hayward} 
\ben
I_H^\prime = \frac{1}{\kappa^2} \int_\gamma \sqrt{\gamma} \eta_0,
\label{1-6}
\een
and a ``tension'' term $I_T$ on surface $Q$
\ben
I_T = -\frac{1}{\kappa^2}\int d\vx dt ~\sqrt{|h|} T.
\label{1-5}
\een
The Lorentzian Hartle-Hawking wavefunctional (which may be thought of as a transition amplitude) is then given by a path integral
\ben
\Psi_{HH}[h_Q,h_\Sigma]= \int {\cal{D}}g_{\mu\nu}~e^{iI}~\delta(g|_Q - h_Q)\delta(g_\Sigma-h_\Sigma).
\label{1-8}
\een
Note that we implicitly impose a boundary condition on $\Sigma$ such that the gravity metric is holographically dual to the quantum state of a CFT on $\Sigma$. In this paper we will then be interested in a semiclassical approximation to $\Psi_{HH}$ which is simply
\ben
\Psi_{HH} \sim e^{iI}|_{on-shell},
\label{1-9}
\een
where total action is 
\ben
I = I_G + I_T + I_H^\prime,
\label{1-7}
\een
and we evaluate it on a classical solution in region $M$. The on-shell action is then maximized with respect to variations of the embedding function $f(t)$ and the holographic path integral complexity is defined as \cite{HHcomplex}
\ben
{\cal{C}} \equiv -(I_G + I_H^\prime),
\label{1-10}
\een
evaluated on the $f(t)$ obtained by the maximization procedure.\footnote{A more careful treatment which includes fluctuations around the saddle point possibly requires use of Picard-Lefshetz theory, e.g. as in \cite{turok}. This issue will not be explored further in this paper.}

For comparisons with our results in later sections let us recall the basic Lorentzian examples \cite{HHcomplex} in pure $AdS_{d+1}$
\ben
ds^2 = \frac{dz^2 - dt^2 + d\vx^2}{z^2}.
\label{1-3}
\een
In both cases, surface $\Sigma$ ($z=\epsilon$) is time-like with trace of extrinsic curvature
\ben
K_\Sigma =d.
\label{1-11}
\een
The induced metric on $Q$, specified by the equation $z = f(t)$, is
\ben
ds^2 = \frac{ -(1-f^\prime(t)^2)dt^2 + d\vx^2}{f^2(t)},
\label{1-12}
\een
and we consider time-like and space-like $Q$'s separately.
%%%%%%%%%%%%%%%
\subsection{Time-like $Q$}
%%%%%%%%%%%%%%%
For a time-like surface $|f^\prime (t)| < 1$. The main object is the trace of the extrinsic curvature on $Q$ given by
\ben
K_Q = \frac{f(t)f^{\prime\prime}(t) - d (1-f^\prime (t)^2)}{(1-f^\prime (t)^2)^{3/2}}.
\label{1-13}
\een
The resulting on-shell action is then further optimized by varying $f(t)$ and this procedure is equivalent to taking $Q$ as a constant meant curvature (CMC) slice of $AdS_{d+1}$ with $K=\frac{d}{d-1}T$.  A solution of this maximization procedure with the condition that $Q$ meets $\Sigma$ at time $t=t_0$ is given by
\ben
f(t) = \frac{d-1}{T} (t-t_0) \sqrt{\frac{T^2}{(d-1)^2}-1}+\epsilon.
\label{1-14}
\een
These surfaces are half-planes parametrized by tension in the range
\ben
-\infty < T < - (d-1),
\label{1-15}
\een
and interpolate between $\Sigma$ for $T\to-(d-1)$ and the null plane for $T\to-\infty$. Then, the hyperbolic angle $\eta_0$ for the embedding surface is given by
\ben
\sinh \eta_0 = \left(\frac{T^2}{(d-1)^2}-1 \right)^{1/2},
\label{1-16}
\een
which leads to the final answer for the HPIC in Lorentzian $AdS_{d+1}$
\ben
{\cal C} = -\frac{(d-1)V_x L_t}{\kappa^2\epsilon^d} - \frac{V_x }{\kappa^2 \epsilon^{d-1}}(\eta_0 - \rm{coth}~\eta_0).
\label{1-17}
\een
In the above, the integration over $t$ is over the range $ -\infty \leq t \leq t_0$ and we can regulate it so that $L_t=t_0-T_0$ for some IR cutoff $T_0$. Then, by simply shifting $T_0\to T_0+t_0$,  we can make the full answer independent of $t_0$ reflecting time-translation invariance of the background.

Last but not the least, the lesson from this holographic complexity is that the optimal surface $Q$ corresponds to $T\to-\infty$ ($\eta_0\to\infty$) i.e, the null plane and matches the intuition from MERA-type TN arguments \cite{Milsted:2018san}.
%%%%%%%%%%%%%%%
\subsection{Space-like $Q$.}
%%%%%%%%%%%%%%%
For a space-like surface $|f^\prime (t)| < 1$ and $K_Q$ is given by
\ben
K_Q = -\frac{f(t)f^{\prime\prime}(t) + d (f^\prime (t)^2)-1)}{(f^\prime (t)^2-1)^{3/2}}.
\label{1-18}
\een
The embedding function for the surface $Q$ meeting $\Sigma$ at time $t=t_0$ is given by
\ben
f(t) = \frac{d-1}{T} (t-t_0) \sqrt{\frac{T^2}{(d-1)^2}+1}+\epsilon.
\label{1-19}
\een
These are again CMC slices of $AdS_{d+1}$ parametrized by the range of the tension between
\ben
-\infty < T < 0,
\label{1-20}
\een
and interpolate between the null sheet at $T\to-\infty$ and the constant time slice of $AdS_{d+1}$ for $T=0$. The hyperbolic angle $\eta_0$ for the embedding surface is now given by
\ben
\sinh \eta_0 = \frac{|T|}{(d-1)}.
\label{1-21}
\een
This leads to the final answer for the HPIC which is almost identical to (\ref{1-17}), but $\eta_0$ is given by (\ref{1-21}) and $\coth\eta_0$ is replaced by $\tanh\eta_0$. Interestingly HPIC is again minimized by light-like $Q$s consistently with MERA-type networks. 

Before we move to our main results, let us point that for the $AdS_{d+1}$, as well as for several other backgrounds considered in \cite{caputa-taka}, similarly to AdS/BCFT \cite{Takayanagi:2011zk}, the saddle point equation obtained by extremizing the Hartle-Hawking wavefunction as a functional of the embedding function $f(t)$ is equivalent to imposing the Neumann boundary conditions on $Q$
\ben
K_{ij}-Kh_{ij}=-Th_{ij}.
\label{NeumannBC}
\een
This also implies (after tracing \eqref{NeumannBC}) that the scalar extrinsic curvature $K_Q$ is constant (CMC). By the virtue of the Hamiltonian constraint, the CMC slices also have constant Ricci scalar as required by the CFT optimization. As we will see below, in some more complicated time-dependent geometries one can find solutions of the maximization which have neither constant Ricci scalar curvature nor are CMC slices. We will also discuss the significance of such examples.

%%%%%%%%%%%%%%%%%%%%%%%%%%%%%%
%%%%%%%%%%%%%%%%%%%%%%%%%%%%%%
\section{$AdS$-Kasner solutions}
\label{adskasner}
%%%%%%%%%%%%%%%%%%%%%%%%%%%%%%
%%%%%%%%%%%%%%%%%%%%%%%%%%%%%%
We now review a family of metrics that will be used to study the HPIC. 
In \cite{dmnt} it was shown that a classical solution of $3+1$ dimensional dilaton gravity with vanishing cosmological constant can be lifted to a solution of $4+1$ dimensional dilaton gravity in the presence of a negative cosmological constant. In this section we will generalize this result to arbitrary number of dimensions and discuss some interesting solutions. 
%%%%%%%%%%%%%%%%%%%
\subsection{The General Result}
%%%%%%%%%%%%%%%%%%%
Consider $d+1$ dimensional dilaton gravity
\ben
S_G = \frac{1}{2\kappa^2}\int_M d^{d+1}x~\sqrt{-g} \left[ R^{d+1} - 2\Lambda - \frac{1}{2} (\nabla \Phi)^2\right] + \frac{1}{\kappa^2} \int_{\partial M}d^dy~\sqrt{|h|} K.
\label{2-1}
\een
We want to look for classical solutions which are of the form
\ben
ds^2 = g_{\mu\nu} dx^\mu dx^\nu = \frac{dz^2 + \tg_{\alpha\beta}(x^\alpha) dx^\alpha dx^\beta }{z^2},
\label{2-2}
\een
where $\alpha,\beta = 1 \cdots d$ while the indices $\mu,\nu = 1 \cdots (d+1)$. Note that the $d$ dimensional metric $\tg_{\alpha\beta}$ is independent of $z$. The Ricci tensor for the metric (\ref{2-2}) can be now easily computed in terms of the Ricci tensor $R^{\tg}_{\alpha\beta}$ of the metric $\tg_{\alpha\beta}$,
\bea
R_{\alpha\beta} =  R^{\tg}_{\alpha\beta} - \frac{d}{z^2} g_{\alpha\beta}, \qquad R_{zz}  =  - \frac{d}{z^2},\qquad R_{z\alpha}  =  0.
\label{2-3}
\eea
Now suppose that
\ben
R^{\tg}_{\alpha\beta} = \frac{1}{2}\partial_\alpha \Phi \partial_\beta \Phi.
\label{2-4}
\een
Using (\ref{2-3}) and (\ref{2-4}) it is straightforward to check that 
\ben
R_{\mu\nu}-\frac{1}{2}g_{\mu\nu} R = \frac{d(d-1)}{2}g_{\mu\nu} + \frac{1}{2} \left[ \partial_\mu \Phi \partial_\nu \Phi - \frac{1}{2} g_{\mu\nu} g^{\mu^\prime \nu^\prime}\partial_{\mu^\prime} \Phi \partial_{\nu^\prime} \Phi \right].
\label{2-5}
\een
Therefore the metric $g_{\mu\nu}$ satisfies the equations of motion of $(d+1)$ dimensional dilaton gravity with a negative cosmological constant given by \eqref{1-2}.
%\ben
%\Lambda = -  \frac{d(d-1)}{2}.
%\label{2-6}
%\een
Note that (\ref{2-4}) is the (trace-reversed) Einstein equation for dilaton gravity with vanishing cosmological constant. This solution can be therefore lifted to a solution in $(d+1)$ dimensions with a negative cosmological constant.

The metric $\tg$ is the boundary metric for the class of metrics (\ref{2-2}). From the point of view of the AdS/CFT correspondence these metrics are somewhat unconventional since $\tg_{\alpha\beta}$ does not  depend on $z$, indicating that in the boundary field theory there is a source for the energy momentum tensor, but a vanishing response. The meaning of this is not very well understood. However this does indicate that the (Heisenberg picture) state of the system is not the vacuum. In this paper we will look for solutions which are translation invariant in the spatial directions of the boundary, i.e. $\tg_{\alpha\beta}(t)$ are functions of time alone.

The last general ingredient that will play an important role in our story is the Hamiltonian constraint. Namely, since we will be interested in the on-shell solutions of \eqref{2-5} written as
\ben
G_{\mu\nu}=-\Lambda g_{\mu\nu}+T^\Phi_{\mu\nu},
\een
we can locally pick a point on an arbitrary slice $Q$ with normal vector $n^\mu$ and project Einstein's equations on its components. In practice this gives the Hamiltonian constraint
\ben\label{HamConstr}
\hat{R}-2\Lambda+\varepsilon(K_{ij}K^{ij}-K^2)= \frac{1}{2}\partial_\mu\Phi\partial^\mu\Phi-\varepsilon(n^\mu\partial_\mu\Phi)^2.
\een
This formula relates the Ricci scalar $\hat{R}$ of $Q$, its extrinsic curvature $K_{ij}$ that has the trace $K$ and the normal components of the dilaton's stress tensor. The normalization $\varepsilon=n^\mu n_\mu$ is equal to $+1$ for time-like and $-1$ for space-like $Q$. In the holographic path integral optimization this will be used as an identity that determines $\hat{R}$. 
%%%%%%%%%%%%%%%%%%%%%%%%%
\subsection{Kasner type solutions}
%%%%%%%%%%%%%%%%%%%%%%%%%
The simplest solutions of this kind are Kasner metrics with homogenous dilaton profile $\Phi =\Phi(t)$. The boundary metric for this class is given by
\ben
ds^2_{b} = \tg_{\alpha\beta}dx^\alpha dx^\beta = -dt^2 + \sum_{a=1}^{(d-1)} t^{2p_a} (dx^a)^2,
\label{2-7}
\een
where $p_a$ are the so-called Kasner exponents. The non-vanishing components of the Ricci tensor for this metric are
\ben
R^{\tg}_{tt}  =  \frac{1}{t^2}\left( \sum_{a=1}^{d-1} p_a - \sum_{a=1}^{d-1} p_a^2 \right), \qquad
R^{\tg}_{aa}  =  \frac{1}{t^2} \left( p_a (\sum_{b=1}^{d-1} p_b -1) t^{2p_a} \right).
\label{2-8}
\een
The equations (\ref{2-4}) can then be solved if either $p_a=0$ for all $a$ or
\begin{align}
\Phi (t)  &= \alpha\log|t|, 
\label{2-9}
\end{align}
together with the constraint for the exponents
\begin{align}
\sum_{a=1}^{d-1} p_a &= 1,~~~~~~~~~\sum_{a=1}^{d-1} p_a^2 = 1-\frac{\alpha^2}{2}.
\label{2-10}
\end{align}
Furthermore, it can be easily seen that (\ref{2-9}) satisfies the equations of motion of the dilaton
\ben
\nabla_{(d+1)}^2 \Phi = 0.
\label{2-10-1}
\een
In particular, one can have symmetric Kasner solutions with all exponents equal
\ben
p_a = \frac{1}{d-1},~~~~ \forall a~~~~~~~\alpha = \left[\frac{2(d-2)}{(d-1)} \right]^{1/2}.
\label{2-11}
\een
When $d=2$ we only have a single $p$, and the only solution to the above constraints becomes $p=1$. Thus from \eqref{2-10}, we have $\alpha = 0$ and hence \eqref{2-9} implies that the dilaton is trivial:  $\Phi(t) = 0$. This is the Milne space-time that is locally diffeomorphic to flat Minkowski space-time. However, if we make the spatial coordinate $x$ compact this is physically different from Minkowski and contains a topological singularity at $t=0$. 

For $d=3$ there are two exponents $p_1,p_2$. For a non-trivial dilaton, there are infinite number of solutions to the equations (\ref{2-10}) parametrized by $\alpha$. These solutions have topological space-like singularities at $t=0$. When the dilaton is constant, the only solution is $p_1=1,p_2=0$ (or $p_1=0, p_2=1$). Again, when the spatial direction is made compact this solution is physically different from Minkowski space-time and singular at $t=0$.
%%%%%%%%%%%%%%%%%%%%%%%%%
\subsection{Solutions with FRW boundaries}
%%%%%%%%%%%%%%%%%%%%%%%%%
Another interesting class of solutions for $d=3$ have boundaries which are FRW universes \cite{adnt}. For instance, when the FRW universe is open a solution of the dilaton gravity system becomes
\bea
ds^2 & = & \frac{1}{z^2}dz^2 + \left[ |\sinh(2t)| \left(-dt^2 + \frac{dr^2}{1+r^2} + r^2 d\Omega_2^2 \right) \right], 
\nonumber \\
\Phi & = & -2\sqrt{3}{\rm tanh}^{-1}(e^t).
\label{2-12}
\eea
The dilaton is now bounded at early and late times. We will not explore these solutions here, but will return to them in the future.
%%%%%%%%%%%%%%%%%%%%%%%%%%%%%%%%%%%%%%%%%%%%%%%%%%
%%%%%%%%%%%%%%%%%%%%%%%%%%%%%%%%%%%%%%%%%%%%%%%%%%
\section{Holographic Path Integral Complexity for Kasner-AdS}
%%%%%%%%%%%%%%%%%%%%%%%%%%%%%%%%%%%%%%%%%%%%%%%%%%
%%%%%%%%%%%%%%%%%%%%%%%%%%%%%%%%%%%%%%%%%%%%%%%%%%
We now use the formalism of section (\ref{sec:review}) to calculate the HPIC for Kasner-AdS solutions of the form (\ref{2-7}) and (\ref{2-9}). As in section (\ref{sec:review}), this requires the calculation of the classical on-shell action
\ben
I = S_G + I_T + I^\prime_H,
\label{3-1}
\een
where $I_T$ and $I^\prime_H$ are defined in (\ref{1-5}) and (\ref{1-6}) and $S_G$ is given by (\ref{2-1}). The manifold $M$ is the region bounded by the surface $z=\epsilon$ denoted by $\Sigma$ and the surface $Q$, which meet at time $t_0$. We will restrict our attention to a situation which is translation invariant in the spatial directions $\vx$ - so that the surface $Q$ is specified by the equation $ z = f(t)$. Again we will perform the calculations separately for time-like and space-like surfaces.

First, the contribution from the bulk term $S_G$ is common for both time-like and space-like $Q$. Using the trace of the equation (\ref{2-5}) the Lagrangian density in $S_G$ is
\ben
R^{(d+1)} - 2\Lambda - \frac{1}{2} (\nabla \Phi)^2 = -2d.
\label{3-2}
\een
Since we have not added an explicit coupling between $Q$ and the dilaton (see also discussion below), this constant result for the bulk integrand will be responsible for the universality of the Hartle-Hawking wavefunction. 

On the other hand, from (\ref{2-2}) and (\ref{2-7}) one has, for this metrics
\ben
\sqrt{-g}=\frac{\sqrt{t^2}}{z^{d+1}},
\label{3-3}
\een
where we have used the first equation of (\ref{2-10}). Thus the contribution from on-shell bulk action is given by
\ben
I_{bulk} = -\frac{d V_x}{\kappa^2}\int_{-T_0}^{t_0} dt \int_\epsilon^{f(t)} dz \frac{\sqrt{t^2}}{z^{d+1}} = \frac{V_x}{\kappa^2}\int_{-T_0}^{t_0} dt \sqrt{t^2} \left( \frac{1}{f(t)^d} - \frac{1}{\epsilon^d} \right).
\label{3-4}
\een 
Here $T_0$ is an infra-red cutoff in the time direction and $V_x$ is the volume of the spatial directions $\vx$. 

The second contribution to the action comes from the Gibbons-Hawking term on $\Sigma$. The normal vector to $\Sigma$ and the volume element coming from the induced metric $h_\Sigma$ on $\Sigma$ are given by 
\ben
n_\Sigma^\mu = -z\delta^{\mu z},~~~~~~\sqrt{-h_\Sigma} = \frac{\sqrt{t^2}}{\epsilon^d}.
\label{3-5}
\een
The trace of the extrinsic curvature of $\Sigma$ is then given by
\ben
K_\Sigma = \nabla_\mu n^\mu_\Sigma = d.
\label{3-6}
\een
This leads to the following contribution from the Gibbons-Hawking term on $\Sigma$
\ben
I^{GH}_\Sigma = \frac{V_x d}{\kappa^2 \epsilon^d} \int_{-T_0}^{t_0} dt \sqrt{t^2}.
\label{3-7}
\een
Contributions from the Gibbons-Hawking term on $Q$, the tension term and the modified Hayward terms need to be evaluated separately for time-like and space-like $Q$ so we do it carefully below.
%%%%%%%%%%%%%%%%%%%%%%%
\subsection{Time-like $Q$}
%%%%%%%%%%%%%%%%%%%%%%%
Time-like surfaces $Q$ are described by functions which obey $|f^\prime(t)|< 1$. Their normal vector is given by
\ben
n^\mu = \frac{z\{f^\prime, {\vec 0},1\}}{\sqrt{1-f^\prime(t)^2}}.
\label{3-8}
\een
The induced metric on the surface is
\ben
h_{ij}dx^idx^j=\frac{1}{f(t)^2}\left[-(1-f^\prime(t)^2)dt^2 +
  \sum_{a=1}^{(d-1)} t^{2p_a}(dx^a)^2 \right],
\label{3-9}
\een
which leads to 
\ben
\sqrt{-h} = \frac{\sqrt{t^2}}{f(t)^d}\sqrt{1-f^\prime(t)^2}.
\label{3-10}
\een
The scalar extrinsic curvature on $Q$ is then given by
\ben
K^t_Q=\frac{t f
  f^{\prime\prime}+ (1-f^\prime (t)^2)(f f^\prime - td)}{t(1-f^\prime(t)^2)^{3/2}},
\label{3-11}
\een
and the superscript $t$ stands for ``time-like''. This leads to the following contribution from the Gibbons-Hawking term on $Q$
\ben
I^{GH}_Q = \frac{1}{\kappa^2}\int_Q \sqrt{-h} (K^t_Q) = 
\frac{V_x}{\kappa^2}\int_{-T_0}^{t_0}\frac{dt \sqrt{t^2}}{f(t)^d} \left[\frac{t f
  f^{\prime\prime}+ (1-f^\prime (t)^2)(f f^\prime - td)}{t(1-f^\prime(t)^2)} \right].
\label{3-12a}
\een
We will then have
\ben
S_G = I_{bulk} + I^{GH}_\Sigma + I^{GH}_Q, 
\label{3-13a}
\een
where the various contributions are given by (\ref{3-4}),(\ref{3-7}) and (\ref{3-12a}). Moreover, the tension term on $Q$ is simply
\ben
I_T = -T\frac{1}{\kappa^2}\int_Q \sqrt{-h}=-T \frac{V_x}{\kappa^2}\int_{-T_0}^{t_0}\frac{dt \sqrt{t^2}}{f(t)^d}
(1-f^\prime(t)^2)^{1/2}.
\label{3-12}
\een

Finally we will need the modified Hayward term. From (\ref{1-4}),(\ref{3-5}) and (\ref{3-8}) the hyperbolic angle $\eta_0$ is given by
\ben
{\rm{cosh}}~ \eta_0 = \frac{1}{\sqrt{1-f^\prime(t_0)^2}},
\label{3-13}
\een
while from (\ref{3-9}) the volume element of the $(d-1)$ dimensional intersection of $\Sigma$ and $Q$ is given by
\ben
\sqrt{-\gamma} = \frac{\sqrt{t_0^2}}{\epsilon^{d-1}},
\label{3-14}
\een
where we have used the fact that $f(t_0) = \epsilon$, and $\sum_a p_a = 1$. The Hayward term then becomes
\ben
I_H^\prime = \frac{1}{\kappa^2}\int_\gamma \sqrt{-\gamma}~\eta_0 = \frac{V_x |t_0|}{\kappa^2 \epsilon^{d-1}} \eta_0.
\label{3-14a}
\een
Adding all together, the total on-shell action as a functional of $f(t)$ becomes
\bea
I & = & \frac{V_x}{\kappa^2}\int_{-T_0}^{t_0} \frac{\sqrt{t^2}}{f(t)^d} \left[ \frac{f^{\prime\prime}(t) f(t)}{1-f^\prime(t)^2} + \frac{f(t)f^\prime(t)}{t} - (d-1) - T\sqrt{1-f^\prime(t)^2} \right] \nonumber \\
&~~~~~~ + & \frac{(d-1)V_x}{\kappa^2 \epsilon^d} \int_{-T_0}^{t_0} dt \sqrt{t^2} +\frac{V_x |t_0|}{\kappa^2 \epsilon^{d-1}}\eta_0.
\label{3-15}
\eea
We will now extremize this on-shell action under variations of $f(t)$ which satisfy $f(t_0) = \epsilon$. In this variational problem the last term in (\ref{3-15}) which is the Hayward term does not contribute to the resulting ``equation of motion''. This extremization equation is given by
\ben
T K_Q = \frac{(d-1)(td-2 f(t) f^\prime (t))}{t(1-f^\prime(t)^2)}+ \frac{2f(t)f^{\prime\prime}(t) ( f(t) f^\prime (t)- (d-1)t)}{t(1-f^\prime(t)^2)^2}.
\label{3-16}
\een
Once we find a solution to (\ref{3-16}) the HPIC is computed by the on-shell value of  
\ben
{\cal C} = -(S_G + I_H^\prime).
\label{3-17}
\een

%%%%%%%%%%%%%%%%%%%%%%%
\subsection{Space-like $Q$}
%%%%%%%%%%%%%%%%%%%%%%%
The calculation for space-like $Q$ is analogous so we just list the relevant formulas. The normal vector to $Q$ is now given by
\ben
n^\mu = \frac{z\{f^\prime, {\vec 0},1\}}{\sqrt{f^\prime(t)^2-1}},
\label{3-18}
\een
whereas the induced volume element on the space-like surface is
\ben
\sqrt{-h} = \frac{\sqrt{t^2}}{f(t)^d}\sqrt{f^\prime(t)^2-1}.
\label{3-19}
\een
The scalar extrinsic curvature on $Q$ is given by
\ben
K^s_Q= -\frac{t f
  f^{\prime\prime}+ (f^\prime(t)^2-1)(td-f f^\prime)}{t(f^\prime(t)^2-1)^{3/2}},
\label{3-20}
\een
where the subscript $s$ denotes that $Q$ is space-like. This leads to the following Gibbons-Hawking term on $Q$
\ben
I_Q = \frac{1}{\kappa^2}\int_Q \sqrt{-h} (K^s_Q) = 
\frac{V_x}{\kappa^2}\int_{-T_0}^{t_0}\frac{dt \sqrt{t^2}}{f(t)^d} \left[-\frac{t f
  f^{\prime\prime}+ (f^\prime(t)^2-1)(td-f f^\prime)}{t(f^\prime(t)^2-1)}\right].
  \label{3-21}
  \een
The contribution from the tension term on $Q$ becomes
\ben
I_T = - \frac{T}{\kappa^2}\int_Q \sqrt{h} = -\frac{TV_x}{\kappa^2}\int_{-T_0}^{t_0}\frac{dt \sqrt{t^2}}{f(t)^d} (f^\prime(t)^2-1)^{1/2}.
\label{3-21a}
\een
In the Hayward term, the hyperbolic angle $\eta_0$ is now given by
\ben
{\rm{sinh}}~ \eta_0 = \frac{1}{\sqrt{f^\prime(t_0)^2-1}},
\label{3-22}
\een
and the volume element of the $(d-1)$ dimensional intersection of $\Sigma$ and $Q$ is the same as in (\ref{3-14}). Adding up all the contributions gives the total on-shell action as a functional of $f(t)$ for space-like slices $Q$
\bea
I & = & \frac{V_x}{\kappa^2}\int_{-T_0}^{t_0} \frac{\sqrt{t^2}}{f(t)^d} \left[- \frac{f^{\prime\prime}(t) f(t)}{f^\prime(t)^2-1} + \frac{f(t)f^\prime(t)}{t} - (d-1) - T\sqrt{f^\prime(t)^2-1} \right] \nonumber \\
&~~~~~~ + & \frac{(d-1)V_x}{\kappa^2 \epsilon^d} \int_{-T_0}^{t_0} dt \sqrt{t^2} + \frac{V_x |t_0|}{\kappa^2 \epsilon^{d-1}}\eta_0.
\label{3-23}
\eea
Note that the first term in the first line of (\ref{3-23}) is exactly the same as the first line of (\ref{3-15}).
This extremization equation is therefore quite similar to (\ref{3-16})
\ben
-T K^s_Q = \frac{(d-1)(2 f(t) f^\prime (t)-td)}{t(f^\prime(t)^2-1)}+ \frac{2f(t)f^{\prime\prime}(t) ( f(t) f^\prime (t)- (d-1)t)}{t(f^\prime(t)^2-1)^2},
\label{3-24}
\een
namely, the right hand side of (\ref{3-24}) is exactly the same as the right hand side of (\ref{3-16}).
%%%%%%%%%%%%%%%%%
\subsection{Universality}
%%%%%%%%%%%%%%%%%
There are several significant features of the on-shell actions (\ref{3-15}) and (\ref{3-23}). 
First, there is no explicit contribution of the dilaton to the
action. The dilaton could have entered through the bulk term, however (\ref{3-2}) shows that the on-shell Lagrangian density does not depend on the dilaton. The effect of the dilaton is entirely through the metric. This fact is well-known for massless bulk fields in AdS/CFT, see e.g. \cite{deharo}.\footnote{We should also point out that the holographic path integral optimization in the presence of scalar fields in the bulk has not been studied in higher dimensions and, similarly to the JT example, may require addition of  extra terms to the bulk action (direct couplings between the scalar and $Q$). Naturally, such extra terms would make the complexity action more sensitive to the properties of $Q$.} Secondly, in the derivation of the various terms in the on-shell action, we only needed to use the first equation of (\ref{2-10}) and never the second equation. Therefore, the on-shell action is completely independent of the parameter $\alpha$ characterizing the dilaton. Thirdly, the Kasner exponents $p_a$ do not enter the expression: the action and the equation of motion (\ref{3-16}) are independent of these exponents as long as $\sum_a p_a = 1$. We therefore conclude that, unlike most of the generic observables, the HPIC action is {\em universal} within the class of Kasner solutions. This universality is a key result of our work. However, as we will see, the properties of the extremal surfaces $Q$ will be very different and dependent on the exponents. This has important consequences on identification of $Q$ with holographic tensor networks.  

%%%%%%%%%%%%%%%%%%%%%%%%%%%%%%%%
\section{Holographic Path Integral Complexity for $d=2$}
%%%%%%%%%%%%%%%%%%%%%%%%%%%%%%%%
To obtain the extremized surface $Q$ and therefore the HPIC we need to solve the equations (\ref{3-16}) and (\ref{3-24}) with the condition that $f(t_0) = \epsilon$ and the function is regular at infinity. Typically it is not easy to obtain analytic solutions, though in principle these equations can be solved numerically. In this section we will obtain analytic solutions for $d=2$ and $d=3$. These solutions and the resulting expressions for the complexity will provide important information about the time dependence of complexity which hopefully generalizes to higher dimensions. Our main conclusion is that the complexity {\em decreases} with time as we approach the singularity. 
%%%%%%%%%%%%%%%%%%%%%%%%%%%%%%%%
\subsection{Solutions for $d=2$}\label{sec:5-1}
%%%%%%%%%%%%%%%%%%%%%%%%%%%%%%%%
For $d=2$ there is only one exponent, $p_1$ and the first equation in (\ref{2-10}) implies $p_1=1$ is the only solution
\ben
ds^2 = \frac{1}{z^2} \left[dz^2 - dt^2 + t^2 dx^2 \right].
\label{4-1}
\een
This means that the dilaton needs to be a constant, $\alpha = 0$.
Locally the boundary metric is diffeomorphic to Minkowski space-time $R^{1,1}$ and the bulk metric is diffeomorphic to Poincare $AdS_3$. This can be seen from the coordinate transformation
\ben
T= t \cosh(x),~~~~~~~X = t\sinh(x),~~~~~Z = z.
\label{4-2}
\een
More precisely, the boundary metric is the Milne universe - the forward light cone of the Minkowski space $(T,X)$. However, when $x$ is a compact coordinate these are not legitimate coordinate transformations, so that the metric (\ref{4-2}) is physically different from Poincare $AdS_3$. The line $t=0$ is then a conical singularity which extends all along the bulk.

We are interested in surfaces which are translationally invariant in the $x$ direction: the compactness of $x$ does not matter. The surface $f(t)$ can be then found by using the coordinate transformation. In \cite{caputa-taka} it was found that the surface $Q$ for pure $AdS_3$ in fact satisfy Neumann boundary conditions for the induced metric, and the most general surface can be written as (see e.g. \cite{Akal:2020wfl})
\ben
(Z-a)^2 + (X-p)^2 - (T-q)^2 = \beta^2,
\label{4-3}
\een
where $a,p,q$ are real parameters. The parameter $\beta$ can be real or purely imaginary. Real $\beta$ corresponds to a time-like $Q$ while a purely imaginary $\beta$ corresponds to a space-like $Q$. The equation (\ref{4-3}) needs to be now written in terms of the Milne coordinates $(z,x,t)$ using (\ref{4-2}). We are considering surfaces which translationally invariant in $x$. This means that we need to set $p=q=0$.
%%%%%%%%%%%%%%%%%%%%%
\subsubsection{Time-like $Q$}
%%%%%%%%%%%%%%%%%%%%%
For time-like $Q$ the surface is given by
\ben
z = f(t) = a \pm \sqrt{t^2 + \beta_t^2},
\label{4-4}
\een
with real and positive $\beta_t$. Reality is obvious as $z$ is real. The demand of positivity comes from the requirement to include the $t = -\infty$ slice for any value of the tension. For negative $\beta_t$ one can see, that the hyperbola $Q$ which intersects $\Sigma$ does so at two finite values of time, thus excluding negative infinity.

\begin{figure}[t!]
    \centering
    \begin{subfigure}[t]{0.43\textwidth}
        \centering
        \includegraphics[width=1\textwidth]{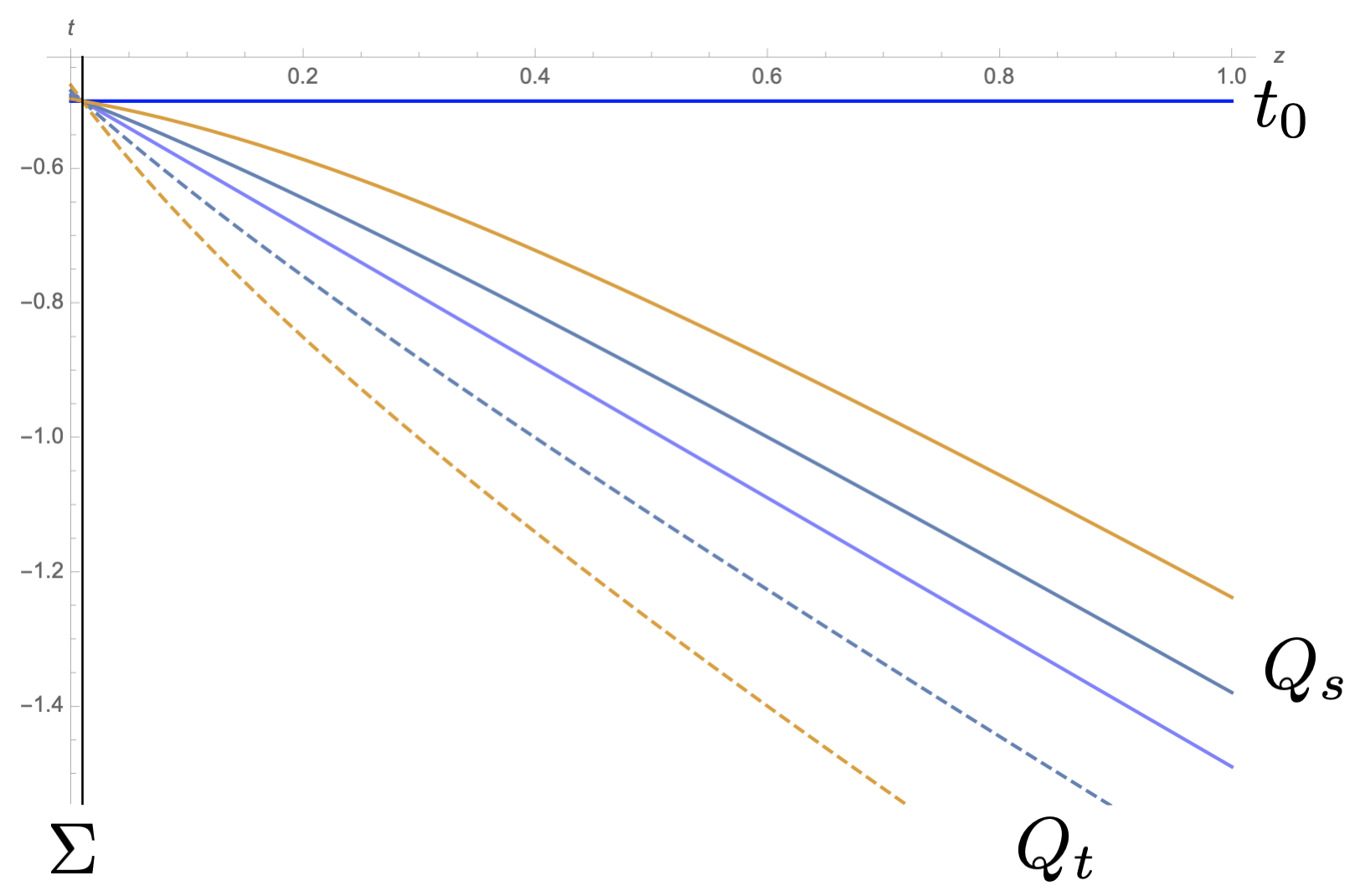}
        \caption{ Surfaces become null for smaller $z$ as $\T$ increases. Time-like $Q_t$ surfaces are plotted for  $\T = 1.3$ (blue) and $1.1$ (orange); and spacelike $Q_s$ for $\T =$ $0.9$ (blue) and $0.3$ (orange). }
    \end{subfigure}%
    ~ 
    \begin{subfigure}[t]{.45\textwidth}
        \centering
        \includegraphics[width=1\textwidth]{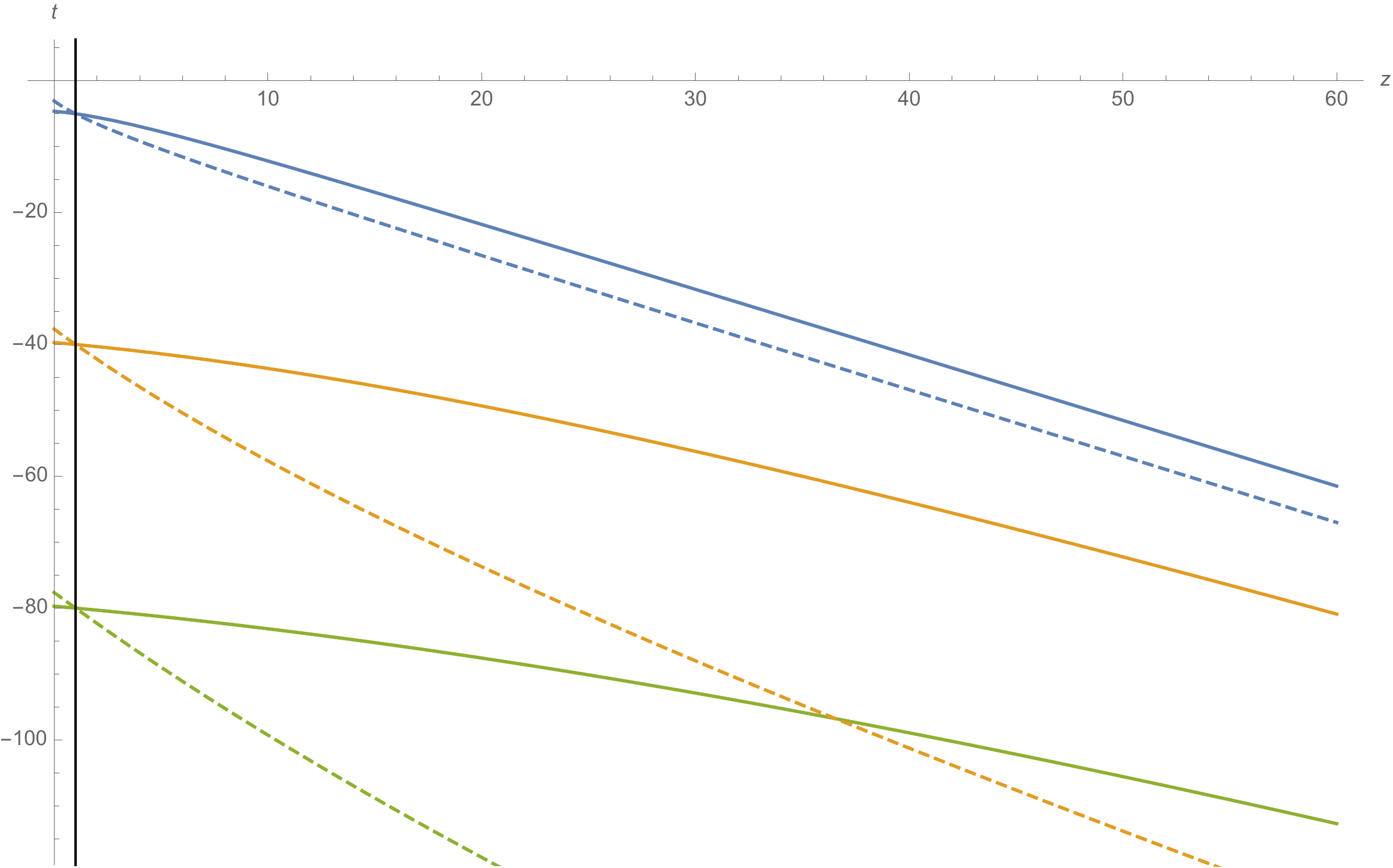}
        \caption{As $|t_0|$ decreases, surfaces become null for smaller $z$. Time-like (dashed) and space-like (solid) $Q$ surfaces  are plotted in the $z-t$ plane, for  $|t_0| = 5$ (blue), $40$ (orange) and $80$ (green).}
    \end{subfigure}
    \caption{Time-like and space-like $Q$ surfaces in the $z-t$ plane (dashed and solid respectively)}
    \label{fig:one}
\end{figure}

It may be now easily verified that this $f(t)$ satisfies the equations of motion (\ref{3-16}) provided
\ben
a = \beta_t T.
\label{4-5}
\een
The condition that $f(t_0) = \epsilon$ then leads to 
\ben
\beta T \pm \sqrt{t_0^2+\beta_t^2} = \epsilon,
\label{4-6}
\een
which determines $\beta_t$ to be
\ben
\beta_t= \frac{\epsilon T \pm \sqrt{t_0^2(T^2-1) +\epsilon^2} }{T^2-1}.
\label{4-6}
\een
If $\beta_t$ is real for arbitrarily small $\epsilon$ (\ref{4-6}) we need $|T| > 1$.
 In the following, we will take $T < 0$ and define $\T = | T |$. Positivity of $\beta_t$ requires us to consider the positive sign in (\ref{4-6}).  Furthermore, we also need to take the positive sign in (\ref{4-4}) to ensure that $Q$ meets $\Sigma$ at some positive value of $z = \epsilon$. Thus the embedding function $f(t)$ for time-like homogeneous $Q$ is given by
\ben
f(t) = -\beta_t \T + \sqrt{t^2+\beta_t^2}.
\label{4-7}
\een
These surfaces are plotted in Fig. \ref{fig:one}. The left panel shows the surfaces for a given value of $|t_0|$ and different value of $\T$. As $\T \rightarrow 1$ the surfaces approach a null surface. The right panel shows the surfaces for three different values of $|t_0|$.

Finally, the trace of the extrinsic curvature on $Q$ is given by
\ben
K^t_Q = \frac{2a}{\beta_t} =2T,
\label{4-8}
\een
so similarly to examples found in \cite{HHcomplex} these are the CMC slices. Then, substituting $d=2$ in (\ref{3-4}), (\ref{3-7}) and (\ref{3-14a}), using (\ref{4-7}) in (\ref{3-12a}) we can now calculate the HPIC
\ben
{\cal C}_t (\T)= - (S_G + I^\prime_H) =- \frac{V_x}{\kappa^2} \left[ \int_{-T_0}^{t_0} dt \sqrt{t^2} \left( \frac{1}{\epsilon^2} + \frac{1-2\T\sqrt{1-f^\prime(t)^2} }{f(t)^2}\right) + \frac{|t_0|}{\epsilon}\eta_0 \right].
\label{4-9}
\een
Again the subscript $t$ on the complexity was put to denote the result for time-like $Q$. The hyperbolic angle  $\eta_0$ is determined by (\ref{3-13}) that leads to
\ben
\sinh(\eta_0) = \frac{|t_0|}{\beta_t}.
\label{4-9a}
\een
We will now examine the behaviour of $\cC_t(\T)$ as a function of $t_0$ for $t_0 < 0$. In particular we are interested in the behaviour near the big crunch for small $t_0$.  The discussion for positive $t_0$ is identical. From the point of view of the dual field theory, $\epsilon$ is a UV cutoff - this means that only $|t_0| > \epsilon$ is meaningful. It is then useful to express everything in terms of the ratio
\ben
x \equiv |t_0|/\epsilon.
\label{4-10}
\een
Physically, since surfaces $Q$ and $\Sigma$ meet at $t_0$, this ratio will measure the distance of $Q$ to the singularity in Kasner geometries and will allow us to scan for its signatures in the holographic optimization.

Using the explicit solution \eqref{4-7} we get
\ben
\frac{\kappa^2}{V_x}\cC_t (\T) = \frac{1}{2} (x^2-(T_0/\epsilon)^2)+\beta_t(x)\T -\log(T_0/\epsilon)-x~ {\rm sinh}^{-1} \left(\frac{x}{\beta_t(x)}\right),
\label{4-11}
\een
where we defined $\beta_t=\epsilon \beta_t(x)$. Similarly to pure $AdS_3$, HPIC \eqref{4-11} is minimal for $\mathcal{T}\to \infty$, so the fully-optimized $Q$ corresponds to MERA-like null slice.

\begin{figure}[t!]
\centering
\includegraphics[width=4.0in]{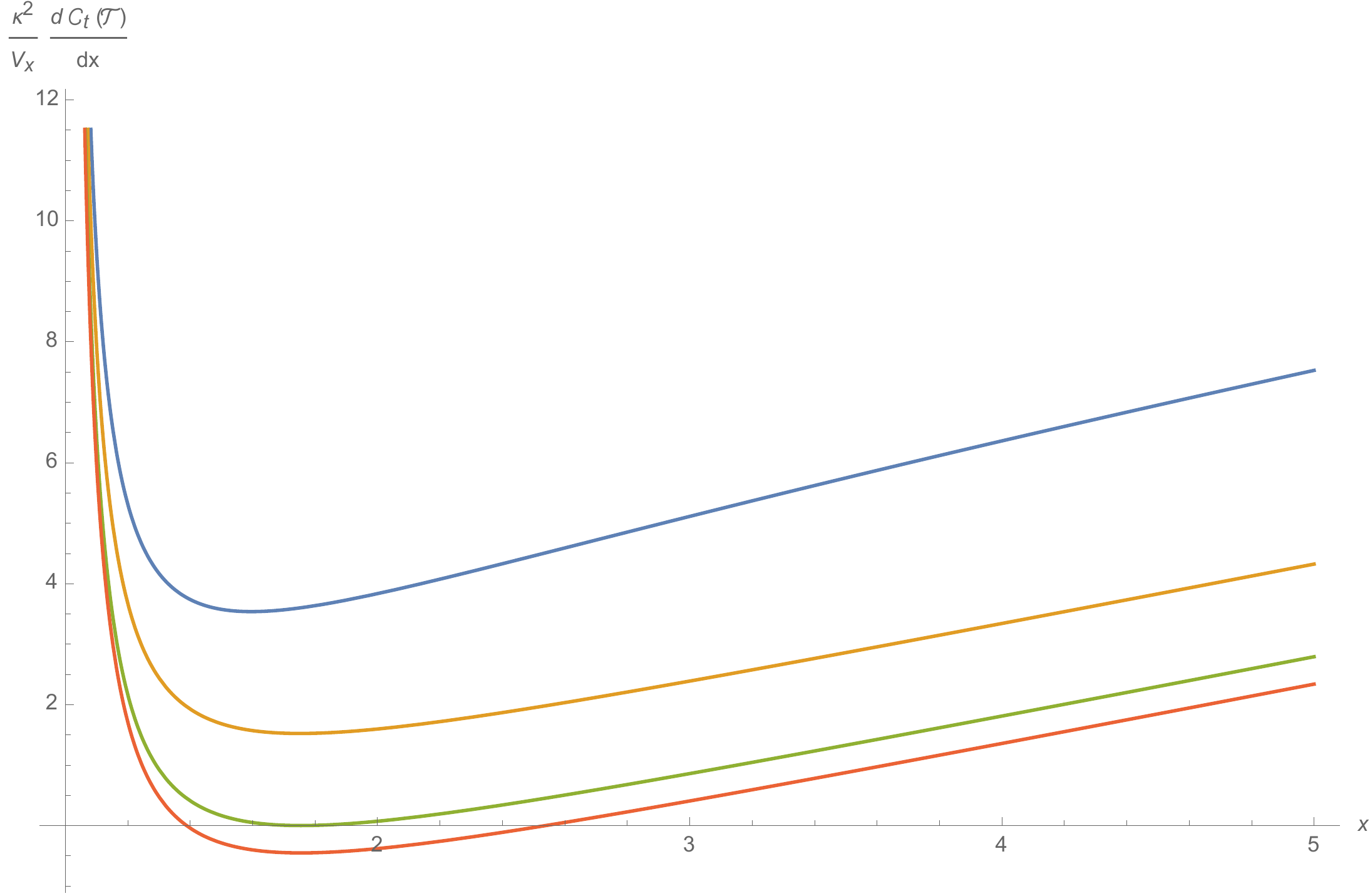}
\caption{Time derivative of the HPIC as a function of $x=|t_0|/\epsilon$ for time-like $Q$ in $d=2$. The curves are for different values of $\T$. From top to bottom $\T = \{1.05,3.00,12.77,20.03\}$. }
\centering
\label{fig:two}
\end{figure}

In order to understand the implications of this result it is better to remove the effect of the IR cutoff $T_0$ and focus on the $x$-derivative of complexity. We have plotted $\frac{d\cC_t (\T)}{dx}$  as a function of $x$ for a wide range of values of $\T = |T|$ in Fig. \ref{fig:two} and verified that this quantity is always positive near $x=1$. Thus the HPIC decreases as we approach the singularity. The derivative is also positive for large enough $x$ and for small enough values of $\T$, it remains positive for all times. However there is a critical value of $\T \sim 12.77$ beyond which there is an interval of $x$ over which $\frac{d\cC_t (\T)}{dx}$ is negative. It is tempting to relate this regime of non-monotonicity of complexity to the \emph{switchback} effect \cite{stankind}. This is an effect associated with precursor operators, which are of the form $U(t)^\dagger W U(t)$. For times much larger than the scrambling time, complexity generically grows, however for earlier times partial cancellations between $U$ and $U^\dagger$ may take place, since the operator $W$ may not have {scrambled} enough. Therefore due to the precursors the complexity may get a negative contribution, making its behaviour non-monotonic. Interestingly in our example as well, the rate is negative only in the small $|t_0|$ regime. It will be very interesting to explore this effect better from the path integral perspective but this will require a better understanding of the CFT state and the nature of its excitations.

At $x=1$, i.e., at the time $|t_0| = \epsilon$, both $\cC_t(\T)$ and $\frac{d\cC_t (\T)}{dx}$ diverge. The leading behaviour of $\cC_t(\T)$ near this time is 
\ben
\cC_t(\T) \sim \frac{1}{2} \left(1 - (T_0/\epsilon)^2\right)+\log \left(\frac{x-1}{2\T} \frac{\epsilon}{T_0} \right).
\label{4-14a}
\een
Therefore, the {\em difference} of complexities for different values of the tension $\T$, i.e., $\cC(\T) - \cC(\T^\prime)$, is finite. This seems to suggest that the physically interesting quantity is also the difference from some reference tension $\T = \T_0$.

It also is interesting to comment that both, the divergence of the complexity and its non-monotonicity as a function of the time $t_0$ stems from the contribution of the modified Hayward term. The quantity $\cC_t^\prime (\T) = -S_G$ is indeed finite at $x=1$ (for a given IR cutoff $T_0$) and its derivative with respect to $x$ is finite and positive for all $x$. However, since the Hayward term is necessary for the consistency of the framework \cite{HHcomplex}, it will be worth to test it in other Lorentzian backgrounds and better understand  its implications for non-equilibrium physics.
%%%%%%%%%%%%%%%%%%%%%%%%%%%
\subsubsection{Space-like $Q$}
%%%%%%%%%%%%%%%%%%%%%%%%%%%
Analogous analysis can be performed for space-like surfaces $Q$. The embedding equation for the surface in this case becomes
\ben
z = f(t) = \beta_s T +\sqrt{t^2-\beta_s^2},
\label{4-19}
\een
where we have considered $T < 0$ and a real and positive $\beta_s$. This surface meets $\Sigma$ at $t=t_0$ so the parameter $\beta_s$ is now fixed to 
\ben
\beta_s = \frac{-\T \epsilon +\sqrt{t_0^2(\T^2+1)-\epsilon^2}}{1+\T^2},
\label{4-20}
\een
where we denoted the norm $\T = |T|$. These slices are plotted in Fig. \ref{fig:one}. The scalar extrinsic curvature on $Q$ is again constant negative $K_Q^s = 2T$ (CMC slices). 

Using these expressions, the path integral complexity for space-like $Q$ becomes
\ben
{\cal C}_s = - (S_G + I^\prime_H) =- \frac{V_x}{\kappa^2} \left[ \int_{-T_0}^{t_0} dt \sqrt{t^2} \left( \frac{1}{\epsilon^2} +\frac{1-2\T\sqrt{f^\prime(t)^2-1}}{f(t)^2} \right) + \frac{|t_0|}{\epsilon}\eta_0 \right],
\label{4-21}
\een
where the hyperbolic angle $\eta_0$ is now given by 
\ben
\cosh\left(\eta_0\right) = \frac{|t_0|}{\beta_s}.
\label{4-22}
\een
As we did for time-like $Q$, we will examine this quantity for $t_0 < 0$ for $|t_0| > \epsilon$. The latter condition is now required for reality of $\beta_s$. In terms of the parameter $x$ defined in (\ref{4-10}) we get
\ben
\frac{\kappa^2}{V_x}\cC_s (\T) = \frac{1}{2} (x^2-(T_0/\epsilon)^2)+\beta_s(x)\T-\log(T_0/\epsilon)-x~ {\rm cosh}^{-1} \left(\frac{x}{\beta_s(x)}\right),
\label{4-23}
\een
where we have defined $\beta_s=\epsilon \beta_s(x)$.  The same as for time-like $Q$, the complexity \eqref{4-23} is minimal for $\mathcal{T}\to \infty$ i.e., the null slice corresponds to the fully-optimized path integral tensor network.

Le us finally focus on the $x$-derivative of \eqref{4-23}. Once again the behaviour of $\frac{\kappa^2}{V_x}\frac{d\cC_s(\T)}{dx}$ versus $x$ for different values of $\T$ for space-like $Q$ is similar to Figure \ref{fig:two}. The time rate of change $\frac{d\cC_s (\T)}{dx}$ is positive near $x=1$ as well as for large $x$. For $\T > 12.66$ there is a range of times $x$ for which the derivative becomes negative.  In all cases, the derivative is positive both near the cutoff singularity at $x=1$ as well as times far from the singularity, large $x$. Moreover, the quantity $\cC_s(\T)$ diverges near $x=1$ where it behaves exactly as in the case of time-like $Q$
\ben
\cC_s(\T) \sim \frac{1}{2} \left( 1 - (T_0/\epsilon)^2\right)+ \log \left(\frac{x-1}{2\T} \frac{\epsilon}{T_0} \right)  + \cdots,
\label{4-14b}
\een
hence the difference $\cC_s(\T)-\cC_s(\T^\prime)$ is finite.

%%%%%%%%%%%%%%%%%%%%%%%%%%%%%%%%
%%%%%%%%%%%%%%%%%%%%%%%%%%%%%%%%
\section{Holographic Path Integral Complexity for $d=3$}
%%%%%%%%%%%%%%%%%%%%%%%%%%%%%%%%
%%%%%%%%%%%%%%%%%%%%%%%%%%%%%%%%
For $d \geq 3$ the solutions of dilaton gravity in the bulk generically have genuine curvature singularities. Unfortunately, the analytic treatment gets more complicated with increasing dimensions. When $d=3$ the solutions with constant dilaton are again locally diffeomorphic to $AdS_4$. This is clear from (\ref{2-10}) since in this case the only solutions are $p_1=1,p_2=0$ or $p_1=0,p_2=1$ which are of course equivalent. By the same token, solutions with time dependent dilaton, i.e. $\alpha \neq 0$ in (\ref{2-9}) are more non-trivial, with curvature singularities at $t=0$. These are parametrized by a single real $\alpha$. Naively, as discussed above, the on-shell actions do not depend on the individual $p_a$, provided they obey (\ref{2-10}). However, we will see below that the story is actually more subtle and the Hamiltonian constraint, that governs the Hartle-Hawking wave functions, must be carefully analyzed to specify the properties of $Q$ as well.

%%%%%%%%%%%%%%%%%%%%%
%%%%%%%%%%%%%%%%%%%%%
\subsection{Space-like solutions for $d=3$ and $T=0$}
%%%%%%%%%%%%%%%%%%%%%
%%%%%%%%%%%%%%%%%%%%%
First of all, we should confess that even in $3d$ we have not been able to find exact solutions of the optimization equations (\ref{3-16}) and (\ref{3-24}) for all $T$. Nevertheless, we can make some modest progress for space-like $Q$ in the limit of zero (and small) tension. Indeed, notice that in $d=3$ for $T=0$ the equation (\ref{3-24}) admits a solution
\ben
f(t) = \sqrt{t^2 - b^2},
\label{4-27}
\een
with real $b$. Requiring $f(t_0) = \epsilon$ one gets
\ben
b = \pm \sqrt{t_0^2 - \epsilon^2}.
\label{4-28}
\een
The reality of $b$ demands $|t_0| > \epsilon$ and we will choose the positive sign in (\ref{4-28}). In what follows, we analyze these surfaces carefully and discuss some of their properties from the angle of the HPIC.

We first evaluate the path integral complexity action. Plugging this solution into (\ref{4-27}) with $T=0$, it is straightforward to see that the contribution from term in the square bracket on the first line of (\ref{3-23}) vanishes, leading to
\ben
S_G = \frac{2V_x}{\kappa^2 \epsilon^3}\int_{-T_0}^{t_0} dt~\sqrt{t^2}.
\label{4-29}
\een
Then, the hyperbolic angle $\eta_0$ in the Hayward term is given by
\ben
\cosh~\eta_0 = \frac{|t_0|}{b}.
\label{4-30}
\een
We will consider the complexity for $t_0 < 0$. Performing the integral in (\ref{4-29}) and using the expression for the Hayward term we get the final form for the complexity
\ben
\frac{\kappa^2 \epsilon}{V_x}\cC = \frac{V_x}{\kappa^2 \epsilon}\left[\frac{1}{2} \left( y^2-(T_0/\epsilon)^2\right) - y \cosh^{-1} \left(\frac{y}{\tb(y)}\right) \right],
\label{4-31}
\een
where we have defined the ratio
\ben
y = \frac{|t_0|}{\epsilon},~~~~~~~~\tb(y) = \frac{b}{\epsilon} =\sqrt{y^2 - 1}.
\label{4-32}
\een
The derivative is now
\ben
\frac{\kappa^2 \epsilon}{V_x}\frac{d\cC}{dy} = \frac{y^3}{y^2-1}-\cosh^{-1} \left(\frac{y}{\sqrt{y^2-1}}\right).
\label{4-33}
\een
It is easy to see that the complexity defined above monotonically increases as we go away from the cutoff singularity at $y=1$. As in $d=2$ the complexity itself is large and negative at the singularity.

\begin{figure}[!h]
\centering
\includegraphics[width=4.0in]{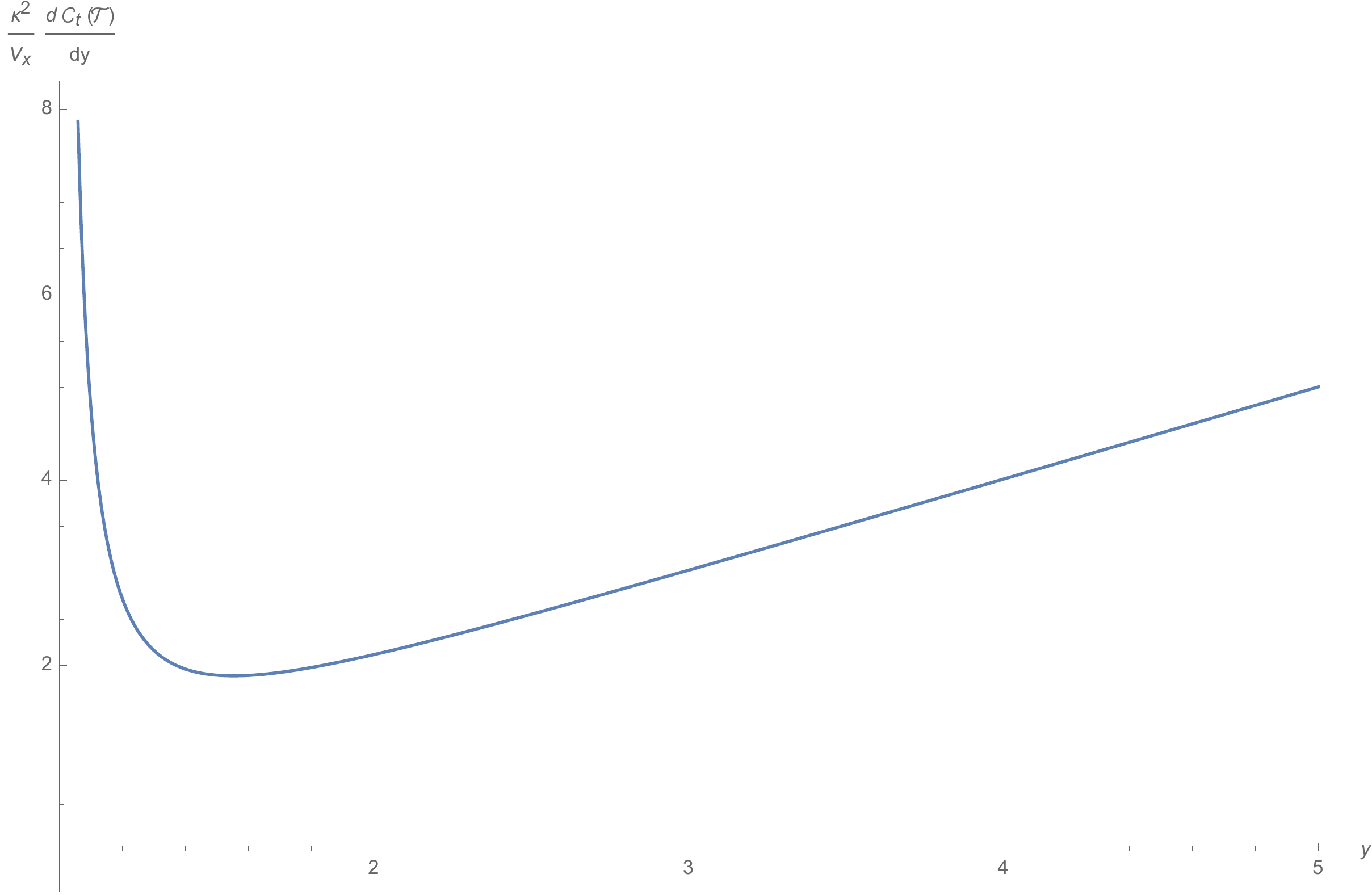}
\caption{Time derivative of the HPIC as a function of $y=|t_0|/\epsilon$ for $T=0$ in $d=3$.}
\centering
\label{fig:three}
\end{figure}
Figure (\ref{fig:three}) is a plot of $\frac{\kappa^2 \epsilon}{V_x}\frac{d\cC}{dy}$ versus $y$. This quantity is clearly positive for all $y$.
Pretty much like the previous cases, the complexity becomes large and negative for $y \rightarrow 1$. 
%%%%%%%%%%%%%%%%%%%%%%%%%%%%%%%%%%%%%%%%%%%%%%%%%%%%%%%%%%%%
\subsubsection{Optimized surfaces and Hamiltonian constraints}
%%%%%%%%%%%%%%%%%%%%%%%%%%%%%%%%%%%%%%%%%%%%%%%%%%%%%%%%%%%%
For the case of $d=2$ the optimized surface is also a CMC surface (with a constant trace of the extrinsic curvature). This was also the case for all the examples considered in \cite{HHcomplex}. However solutions \eqref{4-27}, neither have a constant extrinsic curvature scalar nor solve the Neumann boundary condition. In fact the extrinsic curvature scalar on $Q$ is given by
\ben
K_Q^{d=3} = -\left(\frac{t^2-t_0^2+\epsilon^2}{t_0^2-\epsilon^2}\right)^{1/2}.
\label{4-35}
\een
To get a better picture of the surfaces \eqref{4-27}, let us carefully analyze two bulk solutions in $d=3$ in more detail. The `ambient" metrics and dilatons that solve equations \eqref{2-5} in $d=3$ are
\ben
ds^2_1=\frac{dz^2-dt^2+t^2dx_1^2+dx^2_2}{z^2},\qquad \Phi=const,
\een
and
\ben
ds^2_2=\frac{dz^2-dt^2+t(dx_1^2+dx^2_2)}{z^2},\qquad \Phi=\log|t|.
\een
Therefore, the induced metrics on space-like $Q$ (given by \eqref{4-27}) are different and read
\ben
ds^2_1=\frac{b^2dt^2}{(t^2-b^2)^2}+\frac{t^2dx^2_1+dx^2_2}{t^2-b^2},\qquad ds^2_2=\frac{b^2dt^2}{(t^2-b^2)^2}+\frac{t(dx^2_1+dx^2_2)}{t^2-b^2}.
\een
In particular, their Ricci scalars are
\ben
\hat{R}_1=-6,\qquad\qquad \hat{R}_2=\frac{1}{2}\left(\frac{t^2}{b^2}+\frac{b^2}{t^2}-14\right).\label{RicScalars}
\een
The first result is in fact consistent with the answer required by the path integral optimization (see formula (6.6) in \cite{HHcomplex} for $T=0$) but the second is quite surprizing. Let us first see how this is consistent with the Hamiltonian constraint \eqref{HamConstr}. 

Actually, even though the trace of the extrinsic curvature \eqref{4-35} is the same in both cases, the extrinsic curvature tensors also differ. In the first case, with constant dilaton, we have
\ben
K_{ij}=e^\mu_i e^\nu_j\nabla_\mu n_\nu=\frac{-1}{f^2\sqrt{f'^2-1}}\left(
\begin{array}{ccc}
 f'^2-1+f f'' & 0 & 0 \\
 0 & t \left(t-f f'\right) & 0 \\
 0 & 0 & 1 \\
\end{array}\right),
\een
whereas in the second, with logarithmic dilaton 
\ben
K_{ij}=\frac{-1}{f^2\sqrt{f'^2-1}}\left(
\begin{array}{ccc}
 f'^2-1+f f'' & 0 & 0 \\
 0 & t-\frac{1}{2}ff'  & 0 \\
 0 & 0 & t-\frac{1}{2}ff'  \\
\end{array}\right).
\een
Indeed, we can verify that their traces are the same but, for the solution $Q$ given by \eqref{4-27}, the Neumann boundary condition \eqref{NeumannBC} with $T=0$ is violated. Nevertheless, the relevant combination that appears in the Hamiltonian constraint $K_{ij}K^{ij}-K^2$ vanishes in the case with constant dilaton whereas equals to
\ben
K_{ij}K^{ij}-K^2=\frac{1}{2}-\frac{t^2}{2b^2},\label{CombinationKK}
\een
for the logarithmic dilaton solution. This way, the Hamiltonian constraint \eqref{HamConstr} reproduces $\hat{R}=-6$ for constant $\Phi$ and similarly, for $\Phi=log|t|$ and the right hand side of \eqref{HamConstr}
\ben
\frac{1}{2}\partial_\mu\Phi\partial^\mu\Phi-\varepsilon(n^\mu\partial_\mu\Phi)^2=\frac{t^2}{b^2}+\frac{b^2}{2t^2}-\frac{3}{2},
\een
we recover $\hat{R}_2$ in \eqref{RicScalars}.

A couple of comments are in order at this point. Regarding the first example, the fact that \eqref{4-27} solves the maximization of the Lorentzian Hartle-Hawking wavefunction without Neumann b.c. and still gives constant Ricci scalar on $Q$ (required by CFT optimization) stresses the fact that, in pure gravity, $\hat{R}$ only depends on the particular combination  \eqref{CombinationKK} of the extrinsic geometry data. However, it may still be possible to find a different solution of the Neumann b.c. as well as the optimization equation \eqref{3-24} with $T\neq 0$. We leave this as one of the most important open problems of our analysis in higher dimensions. Regarding the second example, in some sense, it resembles the holographic path integral optimization in JT studied in \cite{HHcomplex}. There, in order for the optimal $Q$ to be the CMC slice, it was important to introduce a direct ``coupling" between the the dilaton and tension $T$ term on $Q$. We expect that a similar mechanism could be introduced here. Moreover, the massless dilaton example is also very special and generalizing this computation of Hartle-Hawking wavefunciton to massive or conformally coupled scalars will be important for better understanding of the influence of bulk matter on the optimization and the Neumann condition on $Q$ in particular.
%%%%%%%%%%%%%%%%%%%%%
%%%%%%%%%%%%%%%%%%%%%
\subsection{Space-like solutions for $d=3$ and small $T$.}
%%%%%%%%%%%%%%%%%%%%%
%%%%%%%%%%%%%%%%%%%%%
A similar solution and analysis can be done to the first order in $T$. Indeed one can check that e.g.
\ben
f(t)=\sqrt{t^2-b^2}+\left(\frac{b}{4}+\frac{a}{\sqrt{t^2-b^2}}\right)T+O(T^2),
\een
also solves \eqref{3-24} to the first order in $T$. Constants $a$ and $b$ can be fixed by the boundary condition $f(t_0)=\epsilon$. Moreover, we can verify that the Neumann b.c. is violated and Ricci scalars in both examples, computed directly from the induced metrics or via the Hamiltonian constraint, become time dependent.  Note that we are expanding (perturbing in $T$) around an extremum. However the status of extrema in Lorentzian path integrals is a subtle issue and needs to be examined much more carefully, presumably with the systematic choice of the integration contours (see e.g. \cite{turok} and references therein). 
%%%%%%%%%%%%%%%%%%%%%%%%%%%%%%%
%%%%%%%%%%%%%%%%%%%%%%%%%%%%%%%
\section{Lessons for Path Integrals in Lorentzian CFTs }
%%%%%%%%%%%%%%%%%%%%%%%%%%%%%%%
%%%%%%%%%%%%%%%%%%%%%%%%%%%%%%%
In this section, we interpret the holographic results derived above from the perspective of path integral optimization in Lorentzian CFTs \cite{HHcomplex} as well as the approach to tensor networks as path integrals in curved spacetimes \cite{Milsted:2018san,Takayanagi:2018pml}. 

As we reviewed in the introduction, the optimization of Euclidean path integrals  in CFTs \cite{caputa-taka} brings us naturally to geometries that solve Liouville equation. The Lorentzian counterpart was discussed in \cite{HHcomplex} and one naturally finds Lorentzian Liouville action playing the role of the path integral complexity. Here, we find analogous story in the non-trivial time-dependent spacetimes. Moreover, in \cite{HHcomplex}, it was argued that surfaces $Q$ can be interpreted as tensor networks preparing states in holographic CFTs. This sharpens the idea \cite{Takayanagi:2018pml} which  identifies metric on $Q$ with an appropriate, optimal way of performing a CFT path integrals. In other words, the curved geometry on which the path integral is computed should be thought of as a general spacetime dependent cutoff, i.e., continuous tensor network. This new way of looking at holographic geometries suggests that holographic spacetime could be interpreted as a collection of different tensor networks (slices) on which CFT path integrals prepare states of dual CFTs.  

With these in mind we will focus on the induced metric on slices Q for the space-like and time-like surfaces in the $AdS_3$-Milne geometry
\ben
ds^2=\frac{dz^2-dt^2+t^2dx^2}{z^2}.
\een
As derived above, slices that maximize Hartle-Hawking wavefunctions in this background solve Neumann boundary condition and have constant mean curvature (CMC) $K=2T$. Below, we will discuss their relation with path integral geometries and their tensor network properties. 

Recall that time-like surfaces $Q$ are specified by the embedding function
\ben
f(t)=\beta_t T+\sqrt{t^2+\beta_t^2},\qquad\qquad \beta_t=\frac{T\epsilon+\sqrt{t^2_0(T^2-1)+\epsilon^2}}{T^2-1}.
\een
This family of CMC slices is parametrized by the tension $-\infty <T<-1$ and interpolate between 
\ben
z=|t|-|t_0|+\epsilon,\qquad T\to-\infty,
\een
and $z=\epsilon$ for $T\to-1$. Note that parameter $\beta_t$ that was fixed by the boundary condition $f(t_0)=\epsilon$ vanishes, ($\beta_t\to0$) as we tune $|t_0|\to\epsilon$. The induced metric on $Q$'s is then
\ben
ds^2=\frac{1}{f^2(t)}\left(-\frac{\beta_t^2}{t^2+\beta_t^2}dt^2+t^2dx^2\right).
\label{QTLS}
\een
We can verify that this two-dimensional geometry has constant positive curvature
\ben
R=2(T^2-1).
\een
This way, analogously to \cite{HHcomplex}, geometry \eqref{QTLS} is a solution of the Lorentzian Liouville equation that should play important role in the optimization for Lorentzian CFTs.\\
Finally, the volume of this time-like slice, that intuitively counts the number of tensors in the network, is given by
\ben
V_t=\int_Q\sqrt{h}=\frac{V_x\beta_t}{\epsilon}.
\een
Since $\beta_t\to 0$ as $|t_0|\to\epsilon$, the number of tensors consistently goes to $0$ indicating decreasing complexity. 

Similarly, we can analyze space-like slices $Q$, specified by
\ben
f(t)=\beta_s T+\sqrt{t^2-\beta_s^2},\qquad \beta_s=\frac{T\epsilon+\sqrt{(1+T^2)t^2_0-\epsilon^2}}{1+T^2},
\een
where again $\beta_s\to0$ as $|t_0|\to\epsilon$. This family, parametrized by $-\infty<T\le 0$, interpolates between
\ben
z=\sqrt{t^2-t^2_0+\epsilon^2},\qquad \text{for}\qquad T=0,
\een
and
\ben
z=|t|-|t_0|+\epsilon,\qquad \text{for}\qquad T\to-\infty.
\een
Their induced metric is
\ben
ds^2=\frac{1}{f(t)^2}\left(\frac{\beta_s^2}{t^2-\beta_s^2}dt^2+t^2dx^2\right).
\een
The Ricci scalar is constant negative
\ben
R=-2(1+T^2),
\een
confirming that the above geometry is a solution equation of motion derived from Lorentzian Liouville action. The volume of this space-like slice is again given by
\ben
V_s=\int_Q\sqrt{h}=\frac{V_x\beta_s}{\epsilon},
\een
and decreases as we tune $|t_0|\to\epsilon$. \\
From the complexity actions we can confirm that the fully optimized Lorentzian tensor networks correspond to $T\to -\infty$ in both solutions, i.e. null surfaces. Similar result was derived in \cite{HHcomplex} for pure Lorentzian $AdS$ geometries where it matches the expectation from MERA-type circuits of path integrals \cite{Milsted:2018san}. Nevertheless, in our context this may be surprising since the MERA geometry for a CFT on the Milne universe could in principle be more complicated.

Let us elaborate more on this interpretation from a slightly different perspective. Namely, in \cite{Milsted:2018san} authors argued that Lorentzian path integral on a thin strip produces a linear map $V$ between Hilbert spaces on its boundaries. More precisely, if we take the metric on the strip to be 
\ben
g_{ij}=\Omega^2(t,x)\left(
\begin{array}{cc}
 -a(t,x)^2+b(t,x)^2 & b(t,x) \\
 b(t,x) & 1 \\
\end{array}
\right),
\een
the infinitesimal linear  map becomes
\ben
\mathcal{V}_t(t)\sim 1-i\epsilon Q_0(t)+i\epsilon Q_1(t),
\een
where
\ben
Q_0(t)=\int dx a(t,x)h(t,x),\qquad Q_1(t)=\int dx b(t,x)p(t,x),
\een
and $h(t,x)$ and $p(t,x)$ correspond to Hamiltonian and momentum densities of a CFT.\\
In this language, our time-like slices \eqref{QTLS} correspond to
\ben
\Omega^2=\frac{t^2}{f(t)^2},\qquad a_t(t)=\frac{\beta_t}{\sqrt{t^2(t^2+\beta_t^2)}},\qquad b=0.
\een
This way the linear map on our geometry corresponds to a Hamiltonian circuit 
\ben
Q^t_0(t)=a_t(t)H.
\een
Analogous derivation can be repeated for space-like $Q$s and they induce linear maps 
\ben
\mathcal{V}_s\sim 1-\epsilon Q^s_0(t),
\een
where
\ben
Q^s_0(t)=a_s(t)H,\qquad a_s(t)=\frac{\beta_s}{\sqrt{t^2(t^2-\beta_s^2)}}.
\een
Firstly, since in both cases $\beta_{t/s}\to 0$ as $\mathcal{T}\to\infty$, therefore $a(t)\to0$ is indicating that the map obtained from hyperbolic or $dS$ slices breaks down and the actual description (TN) is prepared on a MERA-like null sheet \cite{Milsted:2018san}. On the other hand, we can also analyze the behaviour of this map near the singularity as $|t_0|\to \epsilon$. For that is is useful to consider the combination $\epsilon\, a_{t/s}(t)$, that appears in $\mathcal{V}$, at the value of time $t=t_0$ and as a function of $x=|t_0|/\epsilon$. We have explicitly
\ben \label{asFx}
\epsilon\, a_t(x)=\frac{\beta_t(x)}{x\sqrt{x^2+\beta_t(x)^2}},\qquad \epsilon\, a_s(x)=\frac{\beta_s(x)}{x\sqrt{x^2-\beta_t(x)^2}},
\een
with $\beta_t(x)$ and $\beta_s(x)$ defined as before $\beta_{t/s}\equiv \epsilon \beta_{t/s}(x)$. Both expressions vanish for $x=1$, i.e., as we approach the singularity. This suggests that description by the Hamiltonian circuit is not accurate and most likely a different type of Tensor Network should be used to describe the state around that point. This was in fact similar in the case of MERA \cite{Milsted:2018san} that appears as a singular limit of generators derived from path integrals on hyperbolic or de-Sitter geometries. However, in the case of CFTs on singular geometries we expect that actual TN description of states may be more subtle and we leave it as an interesting future direction. 

Interestingly, both expressions in \eqref{asFx} also vanish for $x\to\infty$ and have a maximum that depends on $T$ in the intermediate region of $x>1$. Interpretation of this behaviour and more generally the tension parameter $T$ from the approach \cite{Milsted:2018san} to Tensor Networks is also worth exploring. One can also in principle make a similar analysis for higher-dimensional (e.g. d=3) metrics that we discussed. Even though the MERA-type arguments are less understood, the $T\bar{T}$ ideas \cite{Caputa:2020fbc,Chandra:2021kdv,Jafari:2019qns}, for which the Hamiltonian constraint plays an important role, may be more natural and we hope to return to these holographic tensor networks in the future.

%%%%%%%%%%%%%%%%%%%%%%%%%%%%%%%%%%%%%%%
%%%%%%%%%%%%%%%%%%%%%%%%%%%%%%%%%%%%%%%
\section{Entanglement entropy of 2D CFTs in a time dependent background}
%%%%%%%%%%%%%%%%%%%%%%%%%%%%%%%%%%%%%%%
%%%%%%%%%%%%%%%%%%%%%%%%%%%%%%%%%%%%%%%
In order to probe the dual state in a more standard way, in this section we also consider entanglement entropy in a Lorentzian two-dimensional CFT living in a space that changes with time. In 2d CFT this is a standard computation. A generic situation can be described by the metric
\begin{align}
\label{metric1}
ds^2 &= -dt^2 + a(t)^2 dx^2 = e^{2\varphi} dx^+ dx^-, 
\end{align}
where we introduced
\begin{align}
x^{\pm} &= x \pm \int^t \frac{dt}{a(t) }, \qquad \varphi  = \log a(t). 
\end{align}
In the second equality we have conformally transformed to the flat Minkowski spacetime, which can be ''Euclideanized" to the whole complex plane. Given the map to the plane, correlators maybe efficiently computed with the help of the plane correlators and their transformation under Weyl rescaling
\begin{align}\label{cft-corr}
\vev{O(x_1^+ , x^-_1) O(x_2^+ , x^-_2)}_{e^{2\varphi} dx^+ dx^-} &= \frac{  \vev{O(x_1^+ , x^-_1) O(x_2^+ , x^-_2)}_{ dx^+ dx^-} } { e^{(\varphi(x_1^+ , x^-_1)+\varphi(x_2^+ , x^-_2))\Delta}}.
\end{align}
On the R.H.S, since $x^\pm$ are Minkowski coordinates we know 
 $$\vev{O(x_1^+ , x^-_1) O(x_2^+ , x^-_2)}_{ dx^+ dx^-} =  \left( \frac{x_1^+ - x_2^+}{\epsilon} \right)^{-\Delta} \left( \frac{x_1^- - x_2^-}{\epsilon} \right)^{-\Delta}.$$
Note that we have regularized the coincident point singularity with the UV cutoff $\epsilon$. In general, if the maps $F_\pm(x^\pm)$ take the time-dependent geometry to the plane, then the primary correlator is determined by a further conformal transformation, that yields
\begin{align}\label{cft-corr2}
\vev{O(x_1^+ , x^-_1) O(x_2^+ , x^-_2)}&=  \left( \frac{F_+(x_1^+) -F_+( x_2^+)}{\epsilon\sqrt{ F_+'(x_1^+)F_+'(x_2^+)}} \right)^{-\Delta} \left( \frac{F_-(x_1^-) - F_-(x_2^-)}{\epsilon\sqrt{F_-'(x_1^- )F_-'(x_2^-)}} \right)^{-\Delta}.
\end{align}

The above is a natural Lorentzian generalization of the Euclidean correlators studied recently for inhomogeneous CFTs \cite{Caputa:2020mgb}. The entanglement entropy $S_\ell$, of an interval ($\ell$) at a given time slice is obtained from the reduced density matrix. The reduced density matrix, $\rho_\ell$ is defined by partially tracing out from the full density matrix, the complementary region to the interval. The von Neumann (entanglement) entropy can be computed using the replica trick, which involves considering $\Tr (\rho_\ell^n)$ and then analytically continuing $n \rightarrow 1$. In 2D CFT the replication is implemented naturally by the twist fields which are primary scalars with $\Delta = \frac{c}{12} ( n - 1/n)$ \cite{Calabrese:2009qy}. When the time-dependent spatial geometry is a line then using equation \eqref{cft-corr} and carrying out the above operations results in
\begin{align}
S_\ell &= \frac{c}{6} \log \left( \frac{ a(t)^2 \ell^2}{\epsilon^2}\right).
\end{align}

In case the spatial geometry is compactified with total length $L$, then the map to the plane is given by $F_\pm (x^\pm) = \exp\left(\pm\tfrac{2\pi i}{L} x^{\pm} \right)$. 
When we plug this into eq \eqref{cft-corr2}, \eqref{cft-corr} and implement the analytic continuations we obtain
\begin{align}\label{milne-ent}
S_\ell &= \frac{c}{6} \log \left(  \frac{ L^2 a(t)^2}{\pi^2 \epsilon^2} \sin^2 \frac{ \pi \ell}{L}\right).
\end{align}

For the three dimensional geometry considered in section \ref{sec:5-1} the boundary is the Milne universe, with $x$ direction compactified, $L = 2\pi$ and $a(t) = t$. Therefore we find from the above equation the entanglement entropy as 
\begin{align}
S_\ell &= \frac{c}{3} \log \left(  \frac{ 2 |t|}{ \epsilon} \sin \frac{ \ell}{2}\right).
\end{align}
The regularizing parameter $\epsilon$ being a UV cutoff, it only makes sense to consider times $ t > \epsilon$, and therefore we find that the entanglement entropy decreases as the absolute value of $t$ decreases. The same general result \eqref{milne-ent} can be simply derived holographically using the HRT \cite{Hubeny:2007xt} prescription in Banados geometries \cite{Banados:1998gg}.

%%%%%%%%%%%%%%%%%%%%%%%%%%%%%%%%%%%%%%
\subsection*{Acknowledgements}
%%%%%%%%%%%%%%%%%%%%%%%%%%%%%%%%%%%%%%
We would like to thank Jan Boruch, Nilay Kundu, Tadashi Takayanagi for discussions and comments on the draft. The work of S.R.D. is partially supported by U.S. National Science Foundation grants NSF-PHY/1818878 and NSF/PHY-2111673 and by the Jack and Linda Gill Foundation. D.D. would like to acknowledge the support provided by the Max Planck Partner Group grant MAXPLA/PHY/2018577 and the MATRICS grant {{SERB/PHY/2020334}}. The work of P.C. is supported by NAWA “Polish Returns 2019” and NCN Sonata Bis 9 grants. 

%%%%%%%%%%%%%%%%%%%%%%%%%%%%%%%%%%%%%%
%%%%%%%%%%%%%%%%%%%%%%%%%%%%%%%%%%%%%%

\end{document}